\documentclass[aps,prd,twocolumn,letter]{revtex4}
\usepackage{color}
\usepackage{soul,color}
\usepackage{graphicx}
\usepackage{epsfig}
\usepackage[dvipsnames]{xcolor}

\usepackage{bm}% bold math
\newcommand{\be}{\begin{equation}}
\newcommand{\ee}{\end{equation}}
\newcommand{\ba}{\begin{eqnarray}}
\newcommand{\ea}{\end{eqnarray}}

\begin{document}

\title{Charm Hadrons in pp collisions at LHC energy within a Coalescence plus Fragmentation approach}

\author{Vincenzo Minissale$^{a,b}$, Salvatore Plumari$^{a,b}$ and Vincenzo Greco$^{a,b}$} %\email{}

\affiliation{$^{a}$Department of Physics and Astronomy 'E. Majorana', University of Catania, Via S. Sofia 64, 1-95125 Catania, Italy}
\affiliation{$^{b}$Laboratori Nazionali del Sud, INFN-LNS, Via S. Sofia 62, I-95123 Catania, Italy}

\begin{abstract}
The recent experimental measurements on $pp$ collisions at $\sqrt{s}=5.02 \,\rm TeV$ have shown a very large
abundance of heavy baryon production corresponding to a ratio of $\Lambda_c/D^0 \sim 0.6$, about one
order of magnitude larger than what measured in $e^+e^-$, $ep$ collisions and even in $pp$ collisions at LHC,
but at forward rapidity.  
We apply for the first time to $pp$ collisions a quark coalescence plus fragmentation approach developed for $AA$ collisions, assuming the formation
of Hot QCD matter at finite temperature. An approach that has correctly predicted a $\Lambda_c/D \sim O(1)$  in AA collisions at RHIC energy.
We calculate the heavy baryon/meson ratio and the $p_T$ spectra of charmed hadrons with and 
without strangeness content: $D^{0}$, $D_{s}$, $\Lambda_{c}^{+}$, $\Xi_c$ and
$\Omega_c$ in $pp$ collisions at top LHC energies, finding a satisfactory result for the measured $\Lambda_{c}^{+}/D^0$ 
and the $\Xi_c/D^0$ without any specific tuning of parameters to $pp$ collisions. At variance with other approaches a coalescence approach
predicts also a significant production of $\Omega_c$ such that $\Omega_c/D^0 \sim O(10^{-1})$ .
\end{abstract}



\maketitle

\section{Introduction}
The production of heavy flavored baryons in high energy collisions has become accessible
mainly in the last decade in $e^+e^-$ \cite{Aubert:2006cp}, $e^{\pm}p$ \cite{Abramowicz:2013eja,Abramowicz:2010aa}, 
and in $pp$ collisions at forward rapidity at LHC energy \cite{Aaij:2013mga}.
The hadronization and, in particular, the heavy flavour hadronization  of HQs in $pp$ collisions is 
usually described by the traditional fragmentation mechanism. 
An overall analysis on the charmed hadron production 
have indicated that the charm fragmentation fraction  $f(c \rightarrow D^0)$
is about 0.6, while $f(c \rightarrow D^+) \simeq 0.25$ with a feed-down from resonances such that 
$f(c \rightarrow D^{*}) \simeq 0.25$, while the strange charm meson represents only about a $8-10 \%$ of the production
and for the heavy baryon is estimated a fraction of $f(c \rightarrow \Lambda_c^+) \simeq 0.06$ \cite{Lisovyi:2015uqa}.
Such an heavy baryon production is similar to the one
estimated also by Monte Carlo generators like PYTHIA.

Very recent measurements in $pp$ and $pA$ collisions at top LHC 
in the mid-rapidity region have shown a surprising large production
of the $\Lambda_c$ baryons. Such a production at  low $p_T$ corresponds to $\Lambda_c/D^0 \sim 0.5-0.6$, at variance
with some early observation in $pp$ at large rapidity \cite{Aaij:2013mga}. This is
nearly an order of magnitude larger than what expected with the traditional fragmentation approach.
Furthermore, it has been observed in preliminary results even a significant $\Xi_c$ production with a $\Xi_c/D^0 \sim 0.2-0.3$, while 
the fragmentation approach considers such a production as negligible \cite{Acharya:2021dsq}.

In AA collisions a large $\Lambda_c$ production, again with $\Lambda_c/D^0 \sim O(1)$, 
was predicted about ten years ago assuming
an hadronization via quark coalescence \cite{Oh:2009zj}, an approach that has been able also to
correctly describe several features of the produced hadrons for both light and heavy hadrons
\cite{Greco:2003mm,Minissale:2015zwa,Das:2016llg,Scardina:2017ipo,Plumari:2017ntm,
Dong:2019unq,Wang:2019fcg,Cao:2019iqs,Cho:2019lxb,He:2019vgs,Plumari:2019hzp}. 
Such a prediction was confirmed by the STAR data in $Au+Au$ at 200 AGeV \cite{Adam:2019hpq}
and also by the successive measurements by ALICE in $Pb+Pb$ at 5.02 ATeV \cite{Acharya:2018ckj,Terrevoli:2020fvc}, 
with a similar but somewhat reduced $\Lambda_c/D^0$ ratio,
as again correctly predicted by a coalescence plus fragmentation approach \cite{Plumari:2017ntm}.

In ultra-relativistic nucleus-nucleus collisions  these results have been justified by 
the formation of a deconfined  matter of quarks and gluons (QGP).
The general expectation in elementary $pp$ collision is that 
a QGP is not created,
but  above TeV energy have been observed several features similar to those in $AA$ collisions:
strangeness enhancement \cite{ALICE:2017jyt}, the ridge and large collectivity \cite{Khachatryan:2015lva} and 
enhancement of the baryon to meson ratios \cite{Acharya:2020lrg,Acharya:2020uqi}. 
Theorethical studies of these phenomena in small collision systems have observed that 
hydrodynamics and transport calculations are able to give reasonable description of the $p_T$ spectra and even
two-particle correlations \cite{Weller:2017tsr,Shen:2016zpp,Greif:2017bnr,Sun:2019gxg}, and recently also an approximate
quark number scaling of hadron elliptic flows was shown for these collision systems \cite{Zhao:2019ehg}. 
This would point to the possible formation of a hot QCD matter
at energy density larger than the pseudo-critical one, 
with a lifetime $\tau \approx \, 2 \, \rm fm/c$.

It has to be considered that signatures of hadronization by recombination/coalescence has been spotted even
at FNAL in $\pi^-+A$ by the observation that the $D^-/D^+$ ratio at forward rapidity
gets close to unity against the fragmentation approach predicting a vanishing ratio.
This is also known as "leading particle effect" and is explained as a recombination
of the charm with the valence $d$ quark \cite{Braaten:2002yt}.
This represents a signature that a highly dense quark medium 
favor an hadronization by quark recombination, in this case with valence quark.

An attempt to explain the large charmed baryon production with respect to the
charmed mesons has been proposed in \cite{He:2019tik} assuming a statistical hadronization model (SHM) with the inclusion of a large
set of charm-baryon states beyond the current listings of the Particle Data Group
\cite{Zyla:2020zbs}, and reproducing the $D^0$ and $\Lambda_c$ $p_T-$spectra assuming independent fragmentation 
of charm quarks with the hadronic ratios fixed by the SHM. In such an approach the enhanced feed-down from excited 
charm baryons can account for the large $\Lambda_c /D^0$ ratio at low $p_T$ as measured by ALICE collaboration.
However, the decay of excited charm baryon states proposed have not been seen in the
$e^{+}+e^{-}$ annihilation (Belle \cite{Kniehl:2020szu}).
Another approach able to give an enhancement of the $\Lambda_c$ production
in $pp$ collision is given by a color reconnection mechanism
\cite{Christiansen:2015yqa}, as recently implemented in PYTHIA 8. 
It has been shown that this model is consistent with the CMS result for
the $\Lambda_c/D^0$ ratio \cite{Sirunyan:2019fnc}, but still seems to fail to reproduce a significant production
for higher charmed baryons like $\Xi_c$.

In this Letter, we present results for charm hadron production in $pp$ collisions
at top LHC energy assuming a coalescence plus fragmentation approach occurring
in a bulk matter according to viscous hydro simulations that have
been applied to study the spectra and collectivity in $pp$ collisions \cite{Weller:2017tsr}.
The approach is essentially the same as the one developed for AA collisions 
to study the spectra and the enhanced baryon over meson ratio
for both light and heavy sector \cite{Minissale:2015zwa,Plumari:2017ntm}.

We provide a comprehensive study for the production as a function 
of the transverse momentum of $D^0$, $D_s$, $\Lambda_c^+$, $\Xi_c$ and also the $\Omega_c$
that will be likely been measured in the near future. We also show
the details of the feed-down from resonances. We find a quite satisfying description
of the available data, including an abundant production for $\Xi_c$. The production of these baryons is larger than color  reconnection
in PYTHIA8 \cite{Christiansen:2015yqa} and also to other hadronization approaches including charm states according
to the quark models \cite{He:2019tik}. 
An advantage of such an approach is that it could provide a unified description of charm hadron
production at low and intermediate $p_T$
 in $pp$, $pA$ and $AA$ collisions above the TeV energy scale.

\section{Hybrid hadronization by coalescence and fragmentation}
\label{section:Coal}
The coalescence model was initially proposed as an hadronization
mechanism in heavy ion collisions at RHIC energy 
to explain the $p_T$ spectra and the splitting of elliptic flow of light mesons and baryons
\cite{Greco:2003xt,Fries:2003vb,Greco:2003mm,Fries:2003kq,Molnar:2003ff}.
Subsequent works have been devoted to extended the model to include finite
width to take into account for off-shell effects \cite{Ravagli:2007xx,Ravagli:2008rt,Cassing:2009vt}.
While recently it has been extended to LHC energies to describe mainly the spectra of light hadrons like $\pi, K, p, \phi, \Lambda$ and the baryon to meson ratios at both RHIC and LHC energies
\cite{Minissale:2015zwa}.
For the HF hadron chemistry in $AA$ collisions, it has been
investigated with the coalescence model predicting a large $\Lambda_C/D^0$
\cite{Oh:2009zj,Plumari:2017ntm,Cho:2019lxb,He:2019vgs}. \\

In this section we recall the basic elements of the coalescence model developed
in \cite{Greco:2003mm,Greco:2003vf,Fries:2003kq,Fries:2003vb}
and based on the Wigner formalism. 
The momentum spectrum of hadrons formed by coalescence of quarks can be written as:
\begin{eqnarray}
\label{eq-coal}
\frac{dN_{H}}{dyd^{2}P_{T}}=g_{H} \int \prod^{N_{q}}_{i=1} \frac{d^{3}p_{i}}{(2\pi)^{3}E_{i}} p_{i} \cdot d\sigma_{i}  \; f_{q_i}(x_{i}, p_{i}) \\ 
\times f_{H}(x_{1}...x_{N_{q}}, p_{1}...p_{N_{q}})\, \delta^{(2)} \left(P_{T}-\sum^{n}_{i=1} p_{T,i} \right) \nonumber
\end{eqnarray}
with $g_{H}$ we indicate the statistical factor to form a colorless hadron from quarks and antiquarks with spin 1/2.
The $d\sigma_{i}$ denotes an element of a space-like hypersurface, 
while $f_{q_i}$ are the quark (anti-quark) phase-space distribution functions for i-th quark (anti-quark). 
Finally $f_{H}(x_{1}...x_{N_{q}}, p_{1}...p_{N_{q}})$ is the Wigner function which  
describes the spatial and momentum distribution of quarks in a hadron.
$N_{q}$ is the number of quarks that form the hadron and for $N_{q}=2$ Eq.(\ref{eq-coal})
describes meson formation, while for $N_{q}=3$ the baryon one.
For $D$ mesons the statistical factors $g_{D}=1/36$ gives the probability that two random quarks have the right colour, spin, isospin to match the quantum number of the considered mesons.
For $\Lambda_c$ the statistical factors is $g_{\Lambda_c}=1/108$.
%-----------

Following the Refs. \cite{Oh:2009zj,Plumari:2017ntm,Greco:2003vf}
we adopt for the Wigner distribution function a Gaussian shape in space and momentum, 
\begin{equation}
 f_H(...)=\prod^{N_{q}-1}_{i=1} A_{W}\exp{\Big(-\frac{x_{ri}^2}{\sigma_{ri}^2} - p_{ri}^2 \sigma_{ri}^2\Big)}
\end{equation}
where $N_{q}$ is the number of constituent quarks 
and $A_{W}$ is a normalization constant that it has been fixed to guarantee
that in the limit $p \to 0$ all the charm hadronize by coalescence in a heavy hadron.
This is imposed by requiring that the total coalescence probability gives $\lim_{p \to 0} P^{tot}_{coal}=1$.
It has been shown, by other studies, that the inclusion of missing charm-baryon states \cite{He:2019vgs} 
or the variation of the width of the D meson wave function \cite{Cao:2019iqs,Cho:2019lxb}, can permit 
that all the zero momentum charm quarks can be converted to charmed hadrons.
The 4-vectors for the relative coordinates in space and momentum $x_{ri}$ and $p_{ri}$ are related to the quark coordinate 
in space and momentum by the Jacobian transformations. For mesons the relative coordinates ($x_{r1}$, $p_{r1}$) are given only by 
\begin{eqnarray} 
  x_{r1}=x_{1} - x_{2}, & \, \, & p_{r1}=\frac{m_{2} p_{1}- m_{1} p_{2}}{m_{1}+m_{2}}
\label{Eq:JACOBI1}
\end{eqnarray}
while for baryons we have $x_{r1}$ and $p_{r1}$ and the others two relative coordinates $x_{r2}$, $p_{r2}$ given by
\begin{eqnarray} 
x_{r2}&=&\frac{m_1 x_1 +m_2 x_2}{m_1+m_2}-x_3 \nonumber \\
p_{r2}&=&\frac{m_{3} (p_{1}+p_{2})- (m_{1} +m_{2})p_{3}}{m_{1}+m_{2}+m_{3}}.
\end{eqnarray}
The $\sigma_{ri}$ are the covariant widths, they can be related to the oscillator
frequency $\omega$ by $\sigma_{ri}=1/\sqrt{\mu_i \omega}$ where $\mu_i$ are the reduced masses
\begin{eqnarray} 
\mu_1=\frac{m_1 m_2}{m_1+m_2}, & \, & \mu_2= \frac{(m_1+ m_2)m_3}{m_1+m_2+m_3}.
\end{eqnarray}
In our calculations the masses of light and heavy quarks have been fixed to $m_{u,d}\!=\!300$ MeV, $m_{s}\!=\!380$ MeV, $m_{c}\!=\!1.5$ GeV.The widths of the Wigner function $f_H$ is related to the size of the hadron and in particular to the root mean 
square charge radius of the hadron, $\langle r^2\rangle_{ch}= \sum_{i=1}^{N} Q_i\langle(x_i-X_{cm})^2\rangle$ with $N=2,3$ for mesons and baryons respectively. For mesons, it is given by
\begin{eqnarray} 
\langle r^2\rangle_{ch}&=& \frac{3}{2}  \frac{Q_1 m_2^2+Q_2 m_1^2}{(m_1+m_2)^2} \sigma_r^{2}
\end{eqnarray}
with $Q_i$ the charge of the i-th quark and the center-of-mass 
coordinate calculated as 
\begin{equation}
X_{cm}=\sum_{i=1}^{2} m_i x_i/\sum_{i=1}^2 m_i.
\end{equation}
In a similar way to the mesons, the oscillator frequency and the widths for baryons can be
related to the root mean square charge radius of the corresponding baryons by
\begin{eqnarray} \label{Eq:rBaryon}
\langle r^2\rangle_{ch}&=& \frac{3}{2} \frac{m_2^2 Q_1+m_1^2 Q_2}{(m_1+m_2)^2} \sigma_{r 1}^2  \\ 
&+&\frac{3}{2} \frac{m_3^2 (Q_1+Q_2)+(m_1+m_2)^2 Q_3}{(m_1+m_2+m_3)^2} \sigma_{r 2}^2 \nonumber
\end{eqnarray}
In our approach the Wigner function for the heavy mesons have only one parameter $\sigma_r$ that we fix in order to have their mean square charge radius.
While the Wigner function for Heavy baryons depends on the two widts $\sigma_{r 1}$ and $\sigma_{r 2}$ as shown in
Eq.(\ref{Eq:rBaryon}). However, for baryons there is only one free parameter, because 
the two widths are related by the oscillatory frequency $\omega$ through the reduced
masses by $\sigma_{p i}=\sigma_{r i}^{-1}=1/\sqrt{\mu_{i} \omega}$.
The mean square charge radius of mesons and baryons used in this work have been taken from quark model \cite{Hwang:2001th,Albertus:2003sx}. The corresponding widths for heavy hadron are shown in Table \ref{table:param}.
\begin{table} [ht]
\begin{center}
  \begin{tabular}{l |c c c }
    \hline
    \hline
    Meson &$\langle r^2\rangle_{ch}$ & $\sigma_{p1}$ & $\sigma_{p2}$ \\ 
    $D^{+}=[c \bar{d}]$     & 0.184   & 0.282 & --- \\
    $D_{s}^{+}=[\bar{s}c]$   & 0.083   & 0.404  & --- \\ 
    \hline
    Baryon &$\langle r^2\rangle_{ch}$ & $\sigma_{p1}$ & $\sigma_{p2}$ \\ 
    $\Lambda_c^+ =[udc]$	   & 0.15   & 0.251  & 0.424 \\ 
    $\Xi_c^+ =[usc]$	   & 0.2   & 0.242  & 0.406 \\ 
    $\Omega_c^0 =[ssc]$	   & -0.12   & 0.337  & 0.53 \\ 
\hline
\hline
\end{tabular}
\end{center}
\caption{Mean square charge radius $\langle r^2\rangle_{ch}$ in $fm^2$ and the widths parameters $\sigma_{pi}$ in $GeV$. The mean square charge radius are taken quark model \cite{Hwang:2001th,Albertus:2003sx}.}
\label{table:param}
\end{table}

The multi-dimensional integrals in the coalescence formula are evaluated by using a  
Monte-Carlo method, see \cite{Plumari:2017ntm} for more details.
In these calculations the partons are distributed uniformly in the transverse plane and rapidity $y_{z}$. \\
The hadron momentum spectra from the charm parton fragmentation is given by:
\begin{equation}
\frac{dN_{had}}{d^{2}p_T\,dy}=\sum \int dz \frac{dN_{fragm}}{d^{2}p_T\, dy} \frac{D_{had/c}(z,Q^{2})}{z^{2}} 
\label{Eq:frag}
\end{equation}
$D_{had/c}(z,Q^{2})$ is the fragmentation function and $z=p_{had}/p_{c}$ is the momentum fraction
of heavy quarks transfered to the final heavy hadron while 
$Q^2=(p_{had}/2z)^2$ is the momentum scale for the fragmentation process.

In our calculations we have applied a commonly used fragmentation function for heavy quarks, that is the Peterson fragmentation function \cite{Peterson:1982ak}
$D_{had}(z) \propto 1/[ z [1-z^{-1}-\epsilon_c({1-z})^{-1}]^2 ]$
where $\epsilon_c$ is a free parameter that is determined assuring 
that the shape of the fragmentation function agrees with the experimental data on $p_T$ distributions.
The $\epsilon_c$ parameter has been fixed to $\epsilon_c=0.1$ for both $D^0$ and $\Lambda_c$ that coupled to FONLL $p_T$ distribution correctly describe
the high $p_T$ tail dominated by fragmentation, see Fig. \ref{Fig:D0_Ds_Lc}.
In a similar way done in Ref.\cite{Plumari:2017ntm} for AA collisions we assume that charm quarks that
do not hadronize via coalescence 
are converted to hadrons by fragmentation.
Therefore we can introduce a fragmentation probability given 
by $P_{frag}(p_T)=1- P^{tot}_{coal}(p_T)$, where $P^{tot}_{coal}$ is the total coalescence probability.
The fragmentation fraction that gives the probability that a charm quark fragment in a specific heavy hadron is evaluated according to PYTHIA8 ratios at high $p_T> 10 GeV $ that are similar to the $e^++e^-$ \cite{Lisovyi:2015uqa} apart from an increase of the fraction for $\Lambda_c$ and moderate decrease of the fraction going to $D^0$, as already done in \cite{Plumari:2017ntm}.

\section{Fireball and parton distribution}\label{section:Fireball}

The charm pair production is described by hard process
and it is described by perturbative QCD (pQCD) at NNLO.
Therefore, the starting point to compute the initial heavy quarks spectra
in $pp$ collisions 
at LHC collision energy of
$\sqrt{s}=5.02 \, TeV$ is by pQCD calculation. In our calculation the charm quark spectrum have been taken in accordance 
with the charm distribution in $p + p$ collisions within the Fixed Order + Next-to-Leading Log 
(FONLL), as given in Refs. \cite{Cacciari:2005rk,Cacciari:2012ny}.
In the recent years, we are observing that hydrodynamics models can also be
extended even to extreme situations, for example in small systems like $pA$
collisions, but also even in $pp$ collisions giving reasonable
descriptions of the measured two-particle correlations \cite{Weller:2017tsr,Shen:2016zpp},
suggesting a life time of the fireball,
at these collision energies, of about $\tau \approx 2 \, fm/c$. On the other hand
heavy quarks have a thermalization time that is about $\tau_{th}\approx 5 \, - \, 8 \,fm/c$ which is more than
two times larger than the lifetime estimated for the fireball created in these collisions. 
It is reasonable to assume that the modification of the spectrum due to
the jet quenching mechanism could be negligible, and indeed even in $pA$ measurements
show an $R_{pA} \approx 1$

In our calculation the bulk of particles that we assume is a thermalized system
of gluons and $u,d,s$ quarks and anti-quarks.
The longitudinal momentum distribution is assumed to be boost-invariant in the range $y\in(-0.5,+0.5)$, and is included a radial flow with the following radial profile $\beta_T(r_T)=\beta_{max}\frac{r_T}{R}$,
where $R$ is the transverse radius of the fireball.
Partons at low transverse momentum,  $p_T<2 \,\mbox{GeV} $, are considered thermally distributed
\begin{equation}
\label{quark-distr}
\frac{dN_{q,\bar{q}}}{d^{2}r_{T}\:d^{2}p_{T}} = \frac{g_{q,\bar{q}} \tau m_{T}}{(2\pi)^{3}} \exp \left(-\frac{\gamma_{T}(m_{T}-p_{T}\cdot \beta_{T})}{T} \right) 
\end{equation}
where $m_T=\sqrt{p_T^2+m_{q,\bar{q}}^2}$ is the transverse mass.
The factors $g_{q}=g_{\bar{q}}=6$ are the spin-color degeneracy.
The presence of gluons in the quark-gluon plasma is taken into account 
by converting them to quarks and anti-quark pairs according to the flavour compositions, as assumed in \cite{Biro:1994mp,Greco:2003mm}.
For the bulk properties we fix the parameter according to hydro-dynamical simulations \cite{Weller:2017tsr} 
with $\tau=2.5 \, \mbox{fm}/c$, $R=2 \,\mbox{fm}$ and the temperature of the bulk is  
$T_{C}=165 \rm \, \mbox{MeV}$.
For partons at high transverse momentum, $p_T> 2.5 \, \mbox{GeV}$, we consider the
minijets distribution that can be obtained from pQCD calculations and parametrized with a power law function, see \cite{Plumari:2017ntm, Scardina:2010zz,Liu:2006sf}.

\section{Heavy Hadron transverse momentum spectra and ratio}\label{section:spectra}
In this section, we discuss the coalescence probability and will be shown the results for the transverse momentum spectra
of $D^0$, $D_s$ mesons and for $\Lambda_c$ using the model described in 
previous sections for $pp$ collisions at $\sqrt{s}=5 \, \rm \mbox{TeV}$.

The presence of resonance decay has a significant impact because it gives an important contribution to the ground-state spectra. 
In this study we include ground state hadrons as well as the first excited resonances listed in
Table \ref{tab:charm}, which includes the resonances of $D$, $\Lambda_c$, $\Xi_c$ and $\Omega_c$ baryons
as given by the Particle Data Group \cite{Zyla:2020zbs}.
When we consider the resonance states we take into account a suppression factor given by the Boltzmann probability to populate an excited state of energy $E+\Delta E$, at a temperature $T$. This statistical factor is of the form $[m_{H^*}/m_H]^{3/2} \times \exp {\left(-\Delta E /T\right)}$ with $\Delta E=E_{H^*}-E_H$, where $E_{H^*}=\sqrt{p_T^2+m_{H^*}^2}$ and $m_{H^*}$ is the mass of the resonance, with the same approach already used in \cite{Plumari:2017ntm}. 
Recent experimental analysis techniques have unveiled information about the $\Sigma_c$ spectra and their contribution to the total $\Lambda_c$ yield, which offer a unique possibility to test the hadronization models in detail \cite{Hills:HP2020}.

\begin{table} [ht]
\begin{center}
\begin{tabular}{lcclr}
\hline
Meson & Mass(MeV)  & I (J) & Decay modes   & B.R. \\
\hline 
$D^+ =\bar{d}c$		& 1869 & $\frac{1}{2} \,(0)$	&\\
$D^0 =\bar{u}c$ 	& 1865 & $\frac{1}{2} \,(0)$	&\\
$D_{s}^{+} =\bar{s}c$	& 2011 & $0 \,(0)$		& \\
\hline
Resonances    \\
\hline
$D^{*+} $	& 2010 & $\frac{1}{2} \, (1)$	& $D^0 \pi^+$; $D^+ X$    & $68\%$,$32\%$ \\
$D^{*0} $	& 2007 & $\frac{1}{2} \, (1)$	& $D^0 \pi^0$; $D^0 \gamma$    & $62\%$,$38\%$\\
$D_{s}^{*+} $	& 2112 & $0 \, (1)$		& $D_{s}^+ X$ & $100\%$ \\
\hline
\hline
Baryon &  &  & &\\
\hline 
$\Lambda_c^+ =udc$	& 2286 & $0 \, (\frac{1}{2})$	&\\
$\Xi_c^+ =usc$	& 2467 & $\frac{1}{2} \, (\frac{1}{2})$	&\\
$\Xi_c^0 =dsc$	& 2470 & $\frac{1}{2} \, (\frac{1}{2})$	&\\
$\Omega_c^0 =ssc$	& 2695 & $0 \, (\frac{1}{2})$	&\\
\hline
Resonances &    \\
\hline
$\Lambda_c^+$	& 2595 & $0 \, (\frac{1}{2})$	&$\Lambda_c^+ \pi^+ \pi^-$ & $100\%$ \\
$\Lambda_c^+$	& 2625 & $0 \, (\frac{3}{2})$	&$\Lambda_c^+ \pi^+ \pi^-$ & $100\%$\\
$\Sigma_c^+$	& 2455 & $1 \, (\frac{1}{2})$	&$\Lambda_c^+ \pi$ & $100\%$\\
$\Sigma_c^+$	& 2520 & $1 \, (\frac{3}{2})$	&$\Lambda_c^+ \pi$ & $100\%$\\
$\Xi_c^{'+,0}$	& 2578 & $\frac{1}{2} \, (\frac{1}{2})$	&$\Xi_c^{+,0} \gamma$ & $100\%$\\
$\Xi_c^{+}$	& 2645 & $\frac{1}{2} \, (\frac{3}{2})$	&$\Xi_c^{+} \pi^-$, & $100\%$\\
$\Xi_c^{+}$	& 2790 & $\frac{1}{2} \, (\frac{1}{2})$	&$\Xi_c^{'} \pi$, & $100\%$\\
$\Xi_c^{+}$	& 2815 & $\frac{1}{2} \, (\frac{3}{2})$	&$\Xi_c^{'} \pi$, & $100\%$\\
$\Omega_c^{0}$	& 2770 & $0 \, (\frac{3}{2})$	&$\Omega_c^0 \gamma$, & $100\%$\\
\hline
\end{tabular}
\end{center}
\caption{Ground states of charmed mesons and baryons as well as their first excited states including their decay modes with their corresponding branching ratios as given in Particle
  Data Group \cite{Zyla:2020zbs,Agashe:2014kda}.
\label{tab:charm}}
\end{table}

%%%
\begin{figure}[t]
\centering
\includegraphics[scale=0.31, angle=-90,clip]{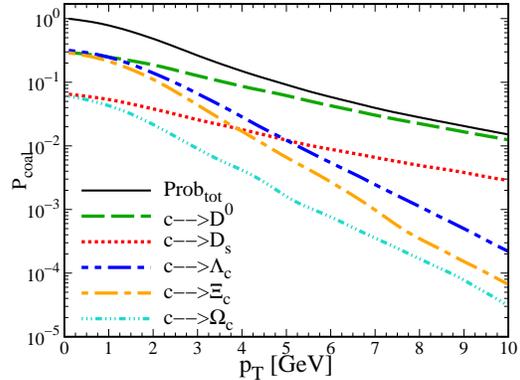}
\caption{
\label{Fig:Pcoalpp}
(Color online) The charm quark coalescence probability as a function of the charm quark $p_T$ for $pp$ collisions at LHC. 
The different lines are the coalescence probabilities to produce the different hadron species. Black solid line is the total coalescence probability.}
\end{figure}
Figure \ref{Fig:Pcoalpp} is shown the coalescence probabilities $P_{coal}$ for a charm quarks to hadronize via coalescence into a specific hadron, as a function of the charm transverse momentum.
As shown $P_{coal}$ is a decreasing function of $p_T$ which means that, at low momentum, 
charm quarks are more probable to hadronize via coalescence with light partons from the thermalized medium, in particular in our model at $p_T \approx 0$ a charm quark can hadronize only by coalescence.
In our modellization a charm quark that cannot hadronize by coalescence hadronizes by fragmentation with a fragmentation probability given by $P_{fragm}=1-P_{coal}$. Therefore at high $p_T$ the fragmentation becomes to be the dominant charm hadronization mechanism and a charm will hadronize according to the different fragmentation fraction into specific final charmed hadron channels, as in Ref. \cite{Lisovyi:2015uqa}.
By comparing the different coalescence probabilities in Fig. \ref{Fig:Pcoalpp} we notice that 
, at low momenta, the coalescence probability for $\Lambda_c$ and $\Xi_c$ are similar than the one for $D^0$ which is a quite peculiar feature of the coalescence mechanism. We expect that this particular characteristic leads to an enhancement of the $\Lambda_c/D^0$ and $\Xi_c/D^0$ ratios. 
\begin{figure}[b]
\centering
\includegraphics[scale=0.32,angle=-90]{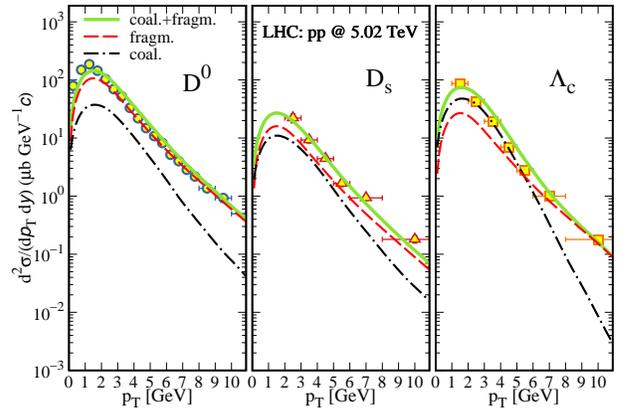}
\caption{
\label{Fig:D0_Ds_Lc}
(Color online) Transverse momentum spectra for $D^0$, $D_s$ mesons and $\Lambda_c$
baryon at mid-rapidity for $pp$ collisions at $\sqrt{s}=5 \, \mbox{TeV}$.
Black dot-dashed and red dashed lines refer to the spectra from only coalescence and
only fragmentation respectively, the green solid line 
is the sum of fragmentation and coalescence. 
Experimental data from \cite{Acharya:2019mgn,Acharya:2020lrg,Acharya:2020uqi}.
}
\end{figure}
In Fig. \ref{Fig:D0_Ds_Lc} we show the $p_T$ spectra of $D^0$ (left panel), $D_s$ (mid panel)
and $\Lambda_c$ (right panel) at mid-rapidity from $pp$ collisions, The total charm cross sections used in this work is
$d\sigma_{c\bar{c}}/dy = 1.0 \mbox{ mb}$. The black dot-dashed line
and the red dashed line refer to the hadron spectra obtained by the contribution
from pure coalescence and pure fragmentation respectively.
We observe that the contribution of fragmentation is the dominant mechanism for the production of $D^0$
in all the $p_T$ range explored and coalescence gives only a few percent of contribution to the total spectrum,
while in $AA$ the contribution is significantly larger and comparable to the fragmentation one \cite{Plumari:2017ntm}.
For the $D_s^+$ spectrum the contribution of both mechanism becomes similar due to the
fact that the fragmentation fraction for $D_s^+$ is quite small, about $8 \%$
of the total heavy hadrons produced, according to Ref. \cite{Lisovyi:2015uqa}. The inclusion of both hadronization mechanisms provide a quite good comparison with the experimental data and the coalescence leads to an enhancement of the $D_s^+$ production.\\
As shown in the last panel on the right of Fig. \ref{Fig:D0_Ds_Lc} the coalescence mechanism is the dominant mechanism for 
the $\Lambda_{c}^{+}$ production for $p_T \lesssim 5 \, \mbox{GeV}$.
This result emerges from the combination of two conditions: 
i) the expected fragmentation fraction into $\Lambda_c^+$ is about 12\% of the total produced heavy hadrons \cite{Lisovyi:2015uqa}, 
ii) the coalescence contribution in the baryon case is dominant with respect to the mesons case
(see \cite{Minissale:2015zwa,Plumari:2017ntm}) because the coalescence mechanism takes quarks that
are already present abundantly in the dense medium created at very high energy even in $pp$ collisions.
\begin{figure}[b]
\centering
\includegraphics[scale=0.32,angle=-90]{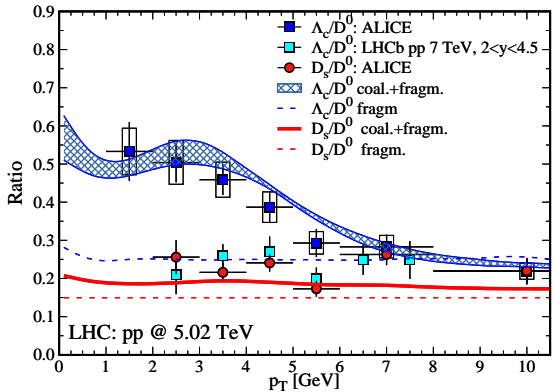}
\caption{
\label{Fig:ratioLcDs}
(Color online) 
$\Lambda_{c}^{+}/D^0$ (blue band) and $D_s^+/D^0$ (red) 
ratios as a function of $p_T$ and at mid-rapidity for 
$pp$ collisions at $\sqrt{s}=5 \, \mbox{TeV}$. Experimental data taken from \cite{Acharya:2019mgn,Acharya:2020lrg,Acharya:2020uqi,Aaij:2013mga}.
Solid and dashed lines refer to the cases with both 
coalescence and fragmentation and to the case with only fragmentation respectively.
}
\end{figure}

In Fig. \ref{Fig:ratioLcDs} we show the results for the $\Lambda_{c}^{+}/D^{0}$ and $D_s^+/D^0$ ratio in comparison with the LHC 
experimental data for $pp$ collisions at $\sqrt{s}=5.02 \, \mbox{TeV}$. The dashed lines show the ratios that comes only from 
fragmentation. 
We have also included the LHCb data for the $\Lambda_{c}^{+}/D^{0}$ at high rapidity, $2<y<4.5$, 
where we expect that the contribution of fragmentation is the dominant one and we can observe a good agreement
with the fragmentation ratio employed (blue dashed lines).
The hybrid approach of coalescence plus fragmentation (solid lines) give a quite good description of the experimental data.
The blue band in the $\Lambda_c/D^0$ ratio corresponds to the uncertainty given by a variation of about 10\% for the Wigner functions widths related to the uncertainties in the mean square charge radius in the Quark Model estimation \cite{Albertus:2003sx,Hwang:2001th}
The coalescence mechanism plays a dominant role in the enhancement of the 
$\Lambda_{c}^{+}$ giving a $\Lambda_{c}^{+}/D^{0} \approx 0.6$ at low $p_T$.
However, the $\Lambda_{c}^{+}/D^{0}$ ratio in $AA$ collision shows a rise and fall behaviour while 
in $pp$ collision the same approach results in a decreasing function of $p_T$, similarly to the experimental data.
At $p_T \lesssim 1\,\rm GeV$ the spectra in Fig.\ref{Fig:D0_Ds_Lc} slightly underestimate the absolute yield for $D^0$
and to some extent also for $\Lambda_c$, hence the rise up of the ratio at low $p_T$ may not be a solid physical result.
The $D_s^+/D^0$ ratio is almost flat in the $p_T$ range explored. 
Comparing the red solid and dashed lines the different relative contribution of coalescence and 
fragmentation for $D_s$ and $D^0$ leads to an enhancement of the ratio $D_s/D^0$ 
of about $20 \%$ in all the range of transverse momentum explored.
\begin{figure}
\centering
\includegraphics[scale=0.31, angle=-90,clip]{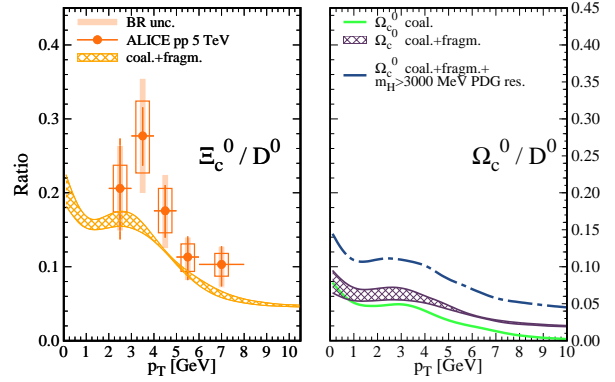}
\caption{
(Color online) Left panel:  
$\Xi_{c}^{0,+}/D^0$ ratios as a function of $p_T$ and at mid-rapidity for 
$pp$ collisions at $\sqrt{s}=5.02 \, \mbox{TeV}$ with both
coalescence and fragmentation, data from \cite{Acharya:2021dsq}.
Right panel: $\Omega_c/D^0$ ratio. For $D^0$ it has been considered coalescence plus fragmentation. For the  $\Omega_c$ are considered cases with only coalescence (green line), coalescence plus fragmentation (purple band), resonances (blue dot-dashed lines). For details see the text.
}\label{Fig:ratioXsiD}
\end{figure}

\begin{figure}
\centering
\includegraphics[scale=0.31,angle=-90]{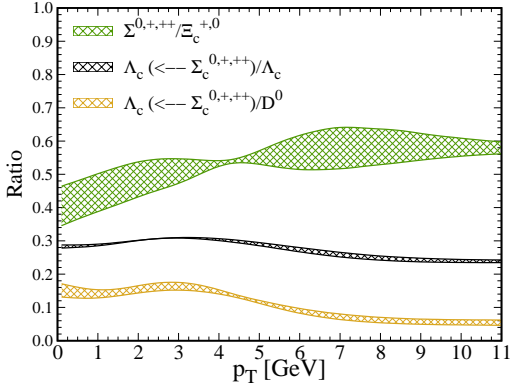}
\caption{
(Color online) Ratios of $\Lambda_c$ that comes from the decay of $\Sigma_c$, $(\Sigma_c \rightarrow \Lambda_{c}^{+})/D^0$ (gold band) and $(\Sigma_c\rightarrow\Lambda_{c}^{+})/\Lambda_{c}$ (black band) as a function of $p_T$ and at mid-rapidity for 
$pp$ collisions at $\sqrt{s}=5.02 \, \mbox{TeV}$. The green band is the $\Sigma_c^{0,+,++}/\Xi_c^{+,0}$ ratio.
}\label{Fig:ratioSigD}
\end{figure}

We have extended this analysis to other single-charmed baryons with 
content of strangeness, such as $\Xi_c$ and $\Omega_c$. The $\Xi_c$ ratios to $D^0$ and $\Lambda_c$ have been 
very recently presented by ALICE collaboration showing
interesting behaviour in $pp$ collision at top LHC energies, in particular, detecting an abundant $\Xi_c$ production
up to $20\%$ of $D^0$, much larger than the expectations in the  fragmentation approach
even including color reconnection \cite{Acharya:2021dsq}.  
In Fig. \ref{Fig:ratioXsiD}, we show the $\Xi_c$ and
$\Omega_c$ to $D^0$ ratios at mid-rapidity in $pp$ collisions at 
$\sqrt{s}=5.02 \, TeV$:  during the revision process of this manuscript 
the $\Xi_c$ data have been published by ALICE collaboration \cite{Acharya:2021dsq} and we included them in the figure. 
In the left panel, we show the $\Xi_c/D^0$ ratio (orange band) including the uncertainties given by the variation of the Wigner function widths. 
Furthermore, we provide some results for the $\Omega_c^0$ baryon that will likely be accessible to
experiments; in the right panel, we show the $\Omega_c^0/D^0$ ratio. For the $D^0$ meson we always consider both coalescence plus fragmentation contribution. For the $\Omega_c^0$ we show three cases, the first considering only the  coalescence contribution (green line), the second with both coalescence plus fragmentation (purple band) and the third where we include
the resonance states present in the PDG in addition to the already considered $\Omega_c^0(2770)$. 
However, currently there are several $\Omega_c^0$ states that have been seen that could give contribution to the
feed-down, but their $J$ values are unknown,
i.e. $\Omega_c(3000)^0$, $\Omega_c(3005)^0$, $\Omega_c(3065)^0$, $\Omega_c(3090)^0$, $\Omega_c(3120)^0$.
To supply an idea of how these states may affect the ratio we have included them by 
assuming for all these resonances $J=3/2$, a kind of average value for their degeneracy (blue dot-dashed line).  
It is interesting that the contribution from fragmentation is subdominant for $\Xi_c$ and $\Omega_c^0$ in these ratios, 
therefore in our model the main contribution comes from a pure coalescence mechanism, so 
this ratio can carry some more relevant information about the hadronization process. 
There are other interesting particle ratios available lately, thanks to the refinement of particle identification technique. 
Shown in Fig. \ref{Fig:ratioSigD} are the ratios between the $\Lambda_c$ particles from the decay of the three $\Sigma_c$ 
states and the total $\Lambda_c$ 
produced (black band) as well as the $D^0$ (gold band). The green band is the ratio between 
the direct $\Sigma_c$ and $\Xi_c$ baryons.
The $(\Sigma_c\rightarrow\Lambda_{c}^{+})/\Lambda_{c}$ ratio is different from the fragmentation observed in $e^+e^-$ for $\Sigma_c$
that is only 10$\%$ of the $\Lambda_c$ fragmentation \cite{Hills:HP2020}. Since these ratios give us information about the contributions from $\Sigma_c$ on the total yield of $\Lambda_c$ they can represents a quite solid test for our approach. 
The preliminary experimental data for the ratios  at $\sqrt{s_{NN}}= 13 TeV$ show a behavior that is quite similar to our calculation 
at least within the current uncertainties. A decreasing behaviour with $p_T$ for the  
$(\Sigma_c\rightarrow\Lambda_{c}^{+})/D^0$ ratio, with a value of about $0.2 \pm 0.05$ at $p_{T} \sim 3 \,$ GeV and $ \sim 0.05$ at $p_{T} \sim 10 \,$ GeV;
while for the $(\Sigma_c\rightarrow\Lambda_{c}^{+})/\Lambda_{c}$ ratio the value is of about $0.4 \pm 0.1$ at $p_{T} \sim 3 \,$ GeV 
and $ \sim 0.25 \pm 0.05$ at $p_{T} \sim 10 \,$ GeV. Furthermore as $\Sigma_c/\Xi_{c}$ ratio exhibits 
an increase with $p_T$, again similar to our modeling,  with a value of about $0.4 \pm 0.15$ at $p_{T} \sim 3 \,$ GeV and $ \sim 0.6 \pm 0.15$ at $p_{T} \sim 10 \,$ GeV.
Hence, it is interesting to note how the hybrid approach of hadronization is able to describe the $p_T$ dependence
as well as the absolute values for these ratios shown in \cite{Hills:HP2020}. 

\section{Conclusion}\label{section:Conclusion}
In this paper we have studied the formation of charmed hadrons, calculating the baryon and meson spectra and their ratios by extending 
a model developed in AA collisions to investigate the production of particles observed in $pp$ collisions at 
top LHC energy.\\
The large baryon over meson ratios with magnitudes much larger than those from 
fragmentation fractions ratio measured in $e^+e^-$ collisions,
clearly indicate a strong violation of the universality of the fragmentation function.
Using our coalescence plus fragmentation model, we have found a good description of the meson and baryon spectra and their ratios.
In particular, in the transition from $AA$ collisions to $pp$ collisions our results naturally describe a change of the baryon over meson ratios 
as a function of the transverse momentum and the enhancement observed in recent experimental data for $\Xi_c/D^0$ \cite{Hills:HP2020} 
ratio along with a prediction for an $\Omega_c/D^0 \approx 0.05-0.1$.

It has to be noticed that once the fireball parameters are given by hydro-dynamical studies, 
the free parameters left in our model are the widths of the wave functions. However these widths are fixed by the charge radii $<r^2>_{ch}$ of charmed hadrons in the constituent quark model. Hence, there are no free parameters specifically
tuned to reproduce the charm hadron production in $pp$ collisions.
Nonetheless, it would be interesting to study possible in-medium modification of the wave function by solving
the pertinent Schrodinger equation for the bound states.
The results obtained seem to suggest that the presence of a hot and dense QCD matter in small collision systems 
permits the recombination of quarks that significantly modify charm baryon production.
In particular, the $\Xi_c/D^0$ \cite{Acharya:2021dsq} and $\Omega_c/D^0$ ratios may shed light on the hadronization mechanism
and provide indications on the differences between the approaches that are currently under development, such as the coalescence approach, the color reconnection mechanism
in the fragmentation approach \cite{Christiansen:2015yqa} and the large feed-down contribution from
higher charmed baryon states \cite{He:2019tik}.

\subsection*{Acknowledgments}
S.P. acknowledge the funding from UniCT under 'Linea di intervento 2' (HQsmall Grant). V.G. acknowledge the funding
from UniCT under 'Linea di intervento 2' (HQCDyn Grant).
This work was supported by the European Union’s Horizon 2020 research and innovation program Strong 2020 under grant agreement No 824093.


\begin{thebibliography}{53}
\expandafter\ifx\csname natexlab\endcsname\relax\def\natexlab#1{#1}\fi
\expandafter\ifx\csname bibnamefont\endcsname\relax
  \def\bibnamefont#1{#1}\fi
\expandafter\ifx\csname bibfnamefont\endcsname\relax
  \def\bibfnamefont#1{#1}\fi
\expandafter\ifx\csname citenamefont\endcsname\relax
  \def\citenamefont#1{#1}\fi
\expandafter\ifx\csname url\endcsname\relax
  \def\url#1{\texttt{#1}}\fi
\expandafter\ifx\csname urlprefix\endcsname\relax\def\urlprefix{URL }\fi
\providecommand{\bibinfo}[2]{#2}
\providecommand{\eprint}[2][]{\url{#2}}

\bibitem[{\citenamefont{Aubert et~al.}(2007)}]{Aubert:2006cp}
\bibinfo{author}{\bibfnamefont{B.}~\bibnamefont{Aubert}} \bibnamefont{et~al.}
  (\bibinfo{collaboration}{BaBar}), \bibinfo{journal}{Phys. Rev. D}
  \textbf{\bibinfo{volume}{75}}, \bibinfo{pages}{012003}
  (\bibinfo{year}{2007}), \eprint{hep-ex/0609004}.

\bibitem[{\citenamefont{Abramowicz et~al.}(2013)}]{Abramowicz:2013eja}
\bibinfo{author}{\bibfnamefont{H.}~\bibnamefont{Abramowicz}}
  \bibnamefont{et~al.} (\bibinfo{collaboration}{ZEUS}), \bibinfo{journal}{JHEP}
  \textbf{\bibinfo{volume}{09}}, \bibinfo{pages}{058} (\bibinfo{year}{2013}).

\bibitem[{\citenamefont{Abramowicz et~al.}(2010)}]{Abramowicz:2010aa}
\bibinfo{author}{\bibfnamefont{H.}~\bibnamefont{Abramowicz}}
  \bibnamefont{et~al.} (\bibinfo{collaboration}{ZEUS}), \bibinfo{journal}{JHEP}
  \textbf{\bibinfo{volume}{11}}, \bibinfo{pages}{009} (\bibinfo{year}{2010}).

\bibitem[{\citenamefont{Aaij et~al.}(2013)}]{Aaij:2013mga}
\bibinfo{author}{\bibfnamefont{R.}~\bibnamefont{Aaij}} \bibnamefont{et~al.}
  (\bibinfo{collaboration}{LHCb}), \bibinfo{journal}{Nucl. Phys. B}
  \textbf{\bibinfo{volume}{871}}, \bibinfo{pages}{1} (\bibinfo{year}{2013}).

\bibitem[{\citenamefont{Lisovyi et~al.}(2016)\citenamefont{Lisovyi, Verbytskyi,
  and Zenaiev}}]{Lisovyi:2015uqa}
\bibinfo{author}{\bibfnamefont{M.}~\bibnamefont{Lisovyi}},
  \bibinfo{author}{\bibfnamefont{A.}~\bibnamefont{Verbytskyi}},
  \bibnamefont{and} \bibinfo{author}{\bibfnamefont{O.}~\bibnamefont{Zenaiev}},
  \bibinfo{journal}{Eur. Phys. J. C} \textbf{\bibinfo{volume}{76}},
  \bibinfo{pages}{397} (\bibinfo{year}{2016}), \eprint{1509.01061}.

%\cite{Acharya:2021dsq}
\bibitem{Acharya:2021dsq}
S.~Acharya \textit{et al.} [ALICE],
%``Measurement of the production cross section of prompt $\Xi^0_{\rm c}$ baryons at midrapidity in pp collisions at $\sqrt{s}$ = 5.02 TeV,''
[arXiv:2105.05616 [nucl-ex]].
%1 citations counted in INSPIRE as of 21 May 2021

\bibitem[{\citenamefont{Hills}(2020)}]{Hills:HP2020}
\bibinfo{author}{\bibfnamefont{C.}~\bibnamefont{Hills}}
  (\bibinfo{collaboration}{ALICE collaboration}), \bibinfo{journal}{{in 10th
  International Conference on Hard and Electromagnetic Probes of High-Energy
  Nuclear Collisions (HP 2020)}, Online, 31 May-5 June 2020}
  (\bibinfo{year}{2020}).

  
\bibitem[{\citenamefont{Oh et~al.}(2009)\citenamefont{Oh, Ko, Lee, and
  Yasui}}]{Oh:2009zj}
\bibinfo{author}{\bibfnamefont{Y.}~\bibnamefont{Oh}},
  \bibinfo{author}{\bibfnamefont{C.~M.} \bibnamefont{Ko}},
  \bibinfo{author}{\bibfnamefont{S.~H.} \bibnamefont{Lee}}, \bibnamefont{and}
  \bibinfo{author}{\bibfnamefont{S.}~\bibnamefont{Yasui}},
  \bibinfo{journal}{Phys. Rev. C} \textbf{\bibinfo{volume}{79}},
  \bibinfo{pages}{044905} (\bibinfo{year}{2009}), \eprint{0901.1382}.

\bibitem[{\citenamefont{Greco et~al.}(2003{\natexlab{a}})\citenamefont{Greco,
  Ko, and Levai}}]{Greco:2003mm}
\bibinfo{author}{\bibfnamefont{V.}~\bibnamefont{Greco}},
  \bibinfo{author}{\bibfnamefont{C.}~\bibnamefont{Ko}}, \bibnamefont{and}
  \bibinfo{author}{\bibfnamefont{P.}~\bibnamefont{Levai}},
  \bibinfo{journal}{Phys.Rev.} \textbf{\bibinfo{volume}{C68}},
  \bibinfo{pages}{034904} (\bibinfo{year}{2003}{\natexlab{a}}),
  \eprint{nucl-th/0305024}.

\bibitem[{\citenamefont{Minissale et~al.}(2015)\citenamefont{Minissale,
  Scardina, and Greco}}]{Minissale:2015zwa}
\bibinfo{author}{\bibfnamefont{V.}~\bibnamefont{Minissale}},
  \bibinfo{author}{\bibfnamefont{F.}~\bibnamefont{Scardina}}, \bibnamefont{and}
  \bibinfo{author}{\bibfnamefont{V.}~\bibnamefont{Greco}},
  \bibinfo{journal}{Phys. Rev. C} \textbf{\bibinfo{volume}{92}},
  \bibinfo{pages}{054904} (\bibinfo{year}{2015}), \eprint{1502.06213}.

\bibitem[{\citenamefont{Das et~al.}(2016)\citenamefont{Das, Torres-Rincon,
  Tolos, Minissale, Scardina, and Greco}}]{Das:2016llg}
\bibinfo{author}{\bibfnamefont{S.~K.} \bibnamefont{Das}},
  \bibinfo{author}{\bibfnamefont{J.~M.} \bibnamefont{Torres-Rincon}},
  \bibinfo{author}{\bibfnamefont{L.}~\bibnamefont{Tolos}},
  \bibinfo{author}{\bibfnamefont{V.}~\bibnamefont{Minissale}},
  \bibinfo{author}{\bibfnamefont{F.}~\bibnamefont{Scardina}}, \bibnamefont{and}
  \bibinfo{author}{\bibfnamefont{V.}~\bibnamefont{Greco}},
  \bibinfo{journal}{Phys. Rev. D} \textbf{\bibinfo{volume}{94}},
  \bibinfo{pages}{114039} (\bibinfo{year}{2016}), \eprint{1604.05666}.

\bibitem[{\citenamefont{Scardina et~al.}(2017)\citenamefont{Scardina, Das,
  Minissale, Plumari, and Greco}}]{Scardina:2017ipo}
\bibinfo{author}{\bibfnamefont{F.}~\bibnamefont{Scardina}},
  \bibinfo{author}{\bibfnamefont{S.~K.} \bibnamefont{Das}},
  \bibinfo{author}{\bibfnamefont{V.}~\bibnamefont{Minissale}},
  \bibinfo{author}{\bibfnamefont{S.}~\bibnamefont{Plumari}}, \bibnamefont{and}
  \bibinfo{author}{\bibfnamefont{V.}~\bibnamefont{Greco}},
  \bibinfo{journal}{Phys. Rev.} \textbf{\bibinfo{volume}{C96}},
  \bibinfo{pages}{044905} (\bibinfo{year}{2017}), \eprint{1707.05452}.

\bibitem[{\citenamefont{Plumari et~al.}(2018)\citenamefont{Plumari, Minissale,
  Das, Coci, and Greco}}]{Plumari:2017ntm}
\bibinfo{author}{\bibfnamefont{S.}~\bibnamefont{Plumari}},
  \bibinfo{author}{\bibfnamefont{V.}~\bibnamefont{Minissale}},
  \bibinfo{author}{\bibfnamefont{S.~K.} \bibnamefont{Das}},
  \bibinfo{author}{\bibfnamefont{G.}~\bibnamefont{Coci}}, \bibnamefont{and}
  \bibinfo{author}{\bibfnamefont{V.}~\bibnamefont{Greco}},
  \bibinfo{journal}{Eur. Phys. J.} \textbf{\bibinfo{volume}{C78}},
  \bibinfo{pages}{348} (\bibinfo{year}{2018}), \eprint{1712.00730}.

\bibitem[{\citenamefont{Dong and Greco}(2019)}]{Dong:2019unq}
\bibinfo{author}{\bibfnamefont{X.}~\bibnamefont{Dong}} \bibnamefont{and}
  \bibinfo{author}{\bibfnamefont{V.}~\bibnamefont{Greco}},
  \bibinfo{journal}{Prog. Part. Nucl. Phys.} \textbf{\bibinfo{volume}{104}},
  \bibinfo{pages}{97} (\bibinfo{year}{2019}).

\bibitem[{\citenamefont{Wang et~al.}(2020)\citenamefont{Wang, Song, Shao, and
  Liang}}]{Wang:2019fcg}
\bibinfo{author}{\bibfnamefont{R.-Q.} \bibnamefont{Wang}},
  \bibinfo{author}{\bibfnamefont{J.}~\bibnamefont{Song}},
  \bibinfo{author}{\bibfnamefont{F.-L.} \bibnamefont{Shao}}, \bibnamefont{and}
  \bibinfo{author}{\bibfnamefont{Z.-T.} \bibnamefont{Liang}},
  \bibinfo{journal}{Phys. Rev. C} \textbf{\bibinfo{volume}{101}},
  \bibinfo{pages}{054903} (\bibinfo{year}{2020}), \eprint{1911.00823}.

\bibitem[{\citenamefont{Cao et~al.}(2020)\citenamefont{Cao, Sun, Li, Liu, Xing,
  Qin, and Ko}}]{Cao:2019iqs}
\bibinfo{author}{\bibfnamefont{S.}~\bibnamefont{Cao}},
  \bibinfo{author}{\bibfnamefont{K.-J.} \bibnamefont{Sun}},
  \bibinfo{author}{\bibfnamefont{S.-Q.} \bibnamefont{Li}},
  \bibinfo{author}{\bibfnamefont{S.~Y.} \bibnamefont{Liu}},
  \bibinfo{author}{\bibfnamefont{W.-J.} \bibnamefont{Xing}},
  \bibinfo{author}{\bibfnamefont{G.-Y.} \bibnamefont{Qin}}, \bibnamefont{and}
  \bibinfo{author}{\bibfnamefont{C.~M.} \bibnamefont{Ko}},
  \bibinfo{journal}{Phys. Lett. B} \textbf{\bibinfo{volume}{807}},
  \bibinfo{pages}{135561} (\bibinfo{year}{2020}), \eprint{1911.00456}.

\bibitem[{\citenamefont{Cho et~al.}(2020)\citenamefont{Cho, Sun, Ko, Lee, and
  Oh}}]{Cho:2019lxb}
\bibinfo{author}{\bibfnamefont{S.}~\bibnamefont{Cho}},
  \bibinfo{author}{\bibfnamefont{K.-J.} \bibnamefont{Sun}},
  \bibinfo{author}{\bibfnamefont{C.~M.} \bibnamefont{Ko}},
  \bibinfo{author}{\bibfnamefont{S.~H.} \bibnamefont{Lee}}, \bibnamefont{and}
  \bibinfo{author}{\bibfnamefont{Y.}~\bibnamefont{Oh}}, \bibinfo{journal}{Phys.
  Rev. C} \textbf{\bibinfo{volume}{101}}, \bibinfo{pages}{024909}
  (\bibinfo{year}{2020}), \eprint{1905.09774}.

\bibitem[{\citenamefont{He and Rapp}(2020)}]{He:2019vgs}
\bibinfo{author}{\bibfnamefont{M.}~\bibnamefont{He}} \bibnamefont{and}
  \bibinfo{author}{\bibfnamefont{R.}~\bibnamefont{Rapp}},
  \bibinfo{journal}{Phys. Rev. Lett.} \textbf{\bibinfo{volume}{124}},
  \bibinfo{pages}{042301} (\bibinfo{year}{2020}), \eprint{1905.09216}.

\bibitem[{\citenamefont{Plumari et~al.}(2020)\citenamefont{Plumari, Coci,
  Minissale, Das, Sun, and Greco}}]{Plumari:2019hzp}
\bibinfo{author}{\bibfnamefont{S.}~\bibnamefont{Plumari}},
  \bibinfo{author}{\bibfnamefont{G.}~\bibnamefont{Coci}},
  \bibinfo{author}{\bibfnamefont{V.}~\bibnamefont{Minissale}},
  \bibinfo{author}{\bibfnamefont{S.~K.} \bibnamefont{Das}},
  \bibinfo{author}{\bibfnamefont{Y.}~\bibnamefont{Sun}}, \bibnamefont{and}
  \bibinfo{author}{\bibfnamefont{V.}~\bibnamefont{Greco}},
  \bibinfo{journal}{Phys. Lett. B} \textbf{\bibinfo{volume}{805}},
  \bibinfo{pages}{135460} (\bibinfo{year}{2020}), \eprint{1912.09350}.

\bibitem[{\citenamefont{Adam et~al.}(2020)}]{Adam:2019hpq}
\bibinfo{author}{\bibfnamefont{J.}~\bibnamefont{Adam}} \bibnamefont{et~al.}
  (\bibinfo{collaboration}{STAR}), \bibinfo{journal}{Phys. Rev. Lett.}
  \textbf{\bibinfo{volume}{124}}, \bibinfo{pages}{172301}
  (\bibinfo{year}{2020}), \eprint{1910.14628}.

\bibitem[{\citenamefont{Acharya et~al.}(2019{\natexlab{a}})}]{Acharya:2018ckj}
\bibinfo{author}{\bibfnamefont{S.}~\bibnamefont{Acharya}} \bibnamefont{et~al.}
  (\bibinfo{collaboration}{ALICE}), \bibinfo{journal}{Phys. Lett. B}
  \textbf{\bibinfo{volume}{793}}, \bibinfo{pages}{212}
  (\bibinfo{year}{2019}{\natexlab{a}}), \eprint{1809.10922}.

\bibitem[{\citenamefont{Terrevoli}(2020)}]{Terrevoli:2020fvc}
\bibinfo{author}{\bibfnamefont{C.}~\bibnamefont{Terrevoli}}
  (\bibinfo{collaboration}{ALICE}), \bibinfo{journal}{J. Phys. Conf. Ser.}
  \textbf{\bibinfo{volume}{1602}}, \bibinfo{pages}{012031}
  (\bibinfo{year}{2020}).

\bibitem[{\citenamefont{Adam et~al.}(2017)}]{ALICE:2017jyt}
\bibinfo{author}{\bibfnamefont{J.}~\bibnamefont{Adam}} \bibnamefont{et~al.}
  (\bibinfo{collaboration}{ALICE}), \bibinfo{journal}{Nature Phys.}
  \textbf{\bibinfo{volume}{13}}, \bibinfo{pages}{535} (\bibinfo{year}{2017}),
  \eprint{1606.07424}.

\bibitem[{\citenamefont{Khachatryan et~al.}(2016)}]{Khachatryan:2015lva}
\bibinfo{author}{\bibfnamefont{V.}~\bibnamefont{Khachatryan}}
  \bibnamefont{et~al.} (\bibinfo{collaboration}{CMS}), \bibinfo{journal}{Phys.
  Rev. Lett.} \textbf{\bibinfo{volume}{116}}, \bibinfo{pages}{172302}
  (\bibinfo{year}{2016}), \eprint{1510.03068}.

\bibitem[{\citenamefont{Acharya et~al.}(2020{\natexlab{a}})}]{Acharya:2020lrg}
\bibinfo{author}{\bibfnamefont{S.}~\bibnamefont{Acharya}} \bibnamefont{et~al.}
  (\bibinfo{collaboration}{ALICE}) (\bibinfo{year}{2020}{\natexlab{a}}),
  \eprint{2011.06079}.

\bibitem[{\citenamefont{Acharya et~al.}(2020{\natexlab{b}})}]{Acharya:2020uqi}
\bibinfo{author}{\bibfnamefont{S.}~\bibnamefont{Acharya}} \bibnamefont{et~al.}
  (\bibinfo{collaboration}{ALICE}) (\bibinfo{year}{2020}{\natexlab{b}}),
  \eprint{2011.06078}.

\bibitem[{\citenamefont{Weller and Romatschke}(2017)}]{Weller:2017tsr}
\bibinfo{author}{\bibfnamefont{R.~D.} \bibnamefont{Weller}} \bibnamefont{and}
  \bibinfo{author}{\bibfnamefont{P.}~\bibnamefont{Romatschke}},
  \bibinfo{journal}{Phys. Lett. B} \textbf{\bibinfo{volume}{774}},
  \bibinfo{pages}{351} (\bibinfo{year}{2017}), \eprint{1701.07145}.

\bibitem[{\citenamefont{Shen et~al.}(2017)\citenamefont{Shen, Paquet, Denicol,
  Jeon, and Gale}}]{Shen:2016zpp}
\bibinfo{author}{\bibfnamefont{C.}~\bibnamefont{Shen}},
  \bibinfo{author}{\bibfnamefont{J.-F.} \bibnamefont{Paquet}},
  \bibinfo{author}{\bibfnamefont{G.~S.} \bibnamefont{Denicol}},
  \bibinfo{author}{\bibfnamefont{S.}~\bibnamefont{Jeon}}, \bibnamefont{and}
  \bibinfo{author}{\bibfnamefont{C.}~\bibnamefont{Gale}},
  \bibinfo{journal}{Phys. Rev. C} \textbf{\bibinfo{volume}{95}},
  \bibinfo{pages}{014906} (\bibinfo{year}{2017}).

\bibitem[{\citenamefont{Greif et~al.}(2017)\citenamefont{Greif, Greiner,
  Schenke, Schlichting, and Xu}}]{Greif:2017bnr}
\bibinfo{author}{\bibfnamefont{M.}~\bibnamefont{Greif}},
  \bibinfo{author}{\bibfnamefont{C.}~\bibnamefont{Greiner}},
  \bibinfo{author}{\bibfnamefont{B.}~\bibnamefont{Schenke}},
  \bibinfo{author}{\bibfnamefont{S.}~\bibnamefont{Schlichting}},
  \bibnamefont{and} \bibinfo{author}{\bibfnamefont{Z.}~\bibnamefont{Xu}},
  \bibinfo{journal}{Phys. Rev. D} \textbf{\bibinfo{volume}{96}},
  \bibinfo{pages}{091504} (\bibinfo{year}{2017}), \eprint{1708.02076}.

\bibitem[{\citenamefont{Sun et~al.}(2020)\citenamefont{Sun, Plumari, and
  Greco}}]{Sun:2019gxg}
\bibinfo{author}{\bibfnamefont{Y.}~\bibnamefont{Sun}},
  \bibinfo{author}{\bibfnamefont{S.}~\bibnamefont{Plumari}}, \bibnamefont{and}
  \bibinfo{author}{\bibfnamefont{V.}~\bibnamefont{Greco}},
  \bibinfo{journal}{Eur. Phys. J. C} \textbf{\bibinfo{volume}{80}},
  \bibinfo{pages}{16} (\bibinfo{year}{2020}), \eprint{1907.11287}.

\bibitem[{\citenamefont{Zhao et~al.}(2020)\citenamefont{Zhao, Ko, Liu, Qin, and
  Song}}]{Zhao:2019ehg}
\bibinfo{author}{\bibfnamefont{W.}~\bibnamefont{Zhao}},
  \bibinfo{author}{\bibfnamefont{C.~M.} \bibnamefont{Ko}},
  \bibinfo{author}{\bibfnamefont{Y.-X.} \bibnamefont{Liu}},
  \bibinfo{author}{\bibfnamefont{G.-Y.} \bibnamefont{Qin}}, \bibnamefont{and}
  \bibinfo{author}{\bibfnamefont{H.}~\bibnamefont{Song}},
  \bibinfo{journal}{Phys. Rev. Lett.} \textbf{\bibinfo{volume}{125}},
  \bibinfo{pages}{072301} (\bibinfo{year}{2020}), \eprint{1911.00826}.
\bibitem[{\citenamefont{Braaten et~al.}(2002)\citenamefont{Braaten, Jia, and
  Mehen}}]{Braaten:2002yt}
\bibinfo{author}{\bibfnamefont{E.}~\bibnamefont{Braaten}},
  \bibinfo{author}{\bibfnamefont{Y.}~\bibnamefont{Jia}}, \bibnamefont{and}
  \bibinfo{author}{\bibfnamefont{T.}~\bibnamefont{Mehen}},
  \bibinfo{journal}{Phys. Rev. Lett.} \textbf{\bibinfo{volume}{89}},
  \bibinfo{pages}{122002} (\bibinfo{year}{2002}), \eprint{hep-ph/0205149}.

\bibitem[{\citenamefont{He and Rapp}(2019)}]{He:2019tik}
\bibinfo{author}{\bibfnamefont{M.}~\bibnamefont{He}} \bibnamefont{and}
  \bibinfo{author}{\bibfnamefont{R.}~\bibnamefont{Rapp}},
  \bibinfo{journal}{Phys. Lett. B} \textbf{\bibinfo{volume}{795}},
  \bibinfo{pages}{117} (\bibinfo{year}{2019}).

\bibitem[{\citenamefont{Zyla et~al.}(2020)}]{Zyla:2020zbs}
\bibinfo{author}{\bibfnamefont{P.}~\bibnamefont{Zyla}} \bibnamefont{et~al.}
  (\bibinfo{collaboration}{Particle Data Group}), \bibinfo{journal}{PTEP}
  \textbf{\bibinfo{volume}{2020}}, \bibinfo{pages}{083C01}
  (\bibinfo{year}{2020}).

\bibitem[{\citenamefont{Kniehl et~al.}(2020)\citenamefont{Kniehl, Kramer,
  Schienbein, and Spiesberger}}]{Kniehl:2020szu}
\bibinfo{author}{\bibfnamefont{B.}~\bibnamefont{Kniehl}},
  \bibinfo{author}{\bibfnamefont{G.}~\bibnamefont{Kramer}},
  \bibinfo{author}{\bibfnamefont{I.}~\bibnamefont{Schienbein}},
  \bibnamefont{and}
  \bibinfo{author}{\bibfnamefont{H.}~\bibnamefont{Spiesberger}},
  \bibinfo{journal}{Phys. Rev. D} \textbf{\bibinfo{volume}{101}},
  \bibinfo{pages}{114021} (\bibinfo{year}{2020}), \eprint{2004.04213}.

\bibitem[{\citenamefont{Christiansen and Skands}(2015)}]{Christiansen:2015yqa}
\bibinfo{author}{\bibfnamefont{J.~R.} \bibnamefont{Christiansen}}
  \bibnamefont{and} \bibinfo{author}{\bibfnamefont{P.~Z.}
  \bibnamefont{Skands}}, \bibinfo{journal}{JHEP} \textbf{\bibinfo{volume}{08}},
  \bibinfo{pages}{003} (\bibinfo{year}{2015}), \eprint{1505.01681}.

\bibitem[{\citenamefont{Sirunyan et~al.}(2020)}]{Sirunyan:2019fnc}
\bibinfo{author}{\bibfnamefont{A.~M.} \bibnamefont{Sirunyan}}
  \bibnamefont{et~al.} (\bibinfo{collaboration}{CMS}), \bibinfo{journal}{Phys.
  Lett. B} \textbf{\bibinfo{volume}{803}}, \bibinfo{pages}{135328}
  (\bibinfo{year}{2020}), \eprint{1906.03322}.

\bibitem[{\citenamefont{Greco et~al.}(2003{\natexlab{b}})\citenamefont{Greco,
  Ko, and Levai}}]{Greco:2003xt}
\bibinfo{author}{\bibfnamefont{V.}~\bibnamefont{Greco}},
  \bibinfo{author}{\bibfnamefont{C.}~\bibnamefont{Ko}}, \bibnamefont{and}
  \bibinfo{author}{\bibfnamefont{P.}~\bibnamefont{Levai}},
  \bibinfo{journal}{Phys.Rev.Lett.} \textbf{\bibinfo{volume}{90}},
  \bibinfo{pages}{202302} (\bibinfo{year}{2003}{\natexlab{b}}),
  \eprint{nucl-th/0301093}.

\bibitem[{\citenamefont{Fries et~al.}(2003{\natexlab{a}})\citenamefont{Fries,
  Muller, Nonaka, and Bass}}]{Fries:2003vb}
\bibinfo{author}{\bibfnamefont{R.}~\bibnamefont{Fries}},
  \bibinfo{author}{\bibfnamefont{B.}~\bibnamefont{Muller}},
  \bibinfo{author}{\bibfnamefont{C.}~\bibnamefont{Nonaka}}, \bibnamefont{and}
  \bibinfo{author}{\bibfnamefont{S.}~\bibnamefont{Bass}},
  \bibinfo{journal}{Phys. Rev. Lett.} \textbf{\bibinfo{volume}{90}},
  \bibinfo{pages}{202303} (\bibinfo{year}{2003}{\natexlab{a}}),
  \eprint{nucl-th/0301087}.

\bibitem[{\citenamefont{Fries et~al.}(2003{\natexlab{b}})\citenamefont{Fries,
  Muller, Nonaka, and Bass}}]{Fries:2003kq}
\bibinfo{author}{\bibfnamefont{R.}~\bibnamefont{Fries}},
  \bibinfo{author}{\bibfnamefont{B.}~\bibnamefont{Muller}},
  \bibinfo{author}{\bibfnamefont{C.}~\bibnamefont{Nonaka}}, \bibnamefont{and}
  \bibinfo{author}{\bibfnamefont{S.}~\bibnamefont{Bass}},
  \bibinfo{journal}{Phys.Rev.} \textbf{\bibinfo{volume}{C68}},
  \bibinfo{pages}{044902} (\bibinfo{year}{2003}{\natexlab{b}}),
  \eprint{nucl-th/0306027}.

\bibitem[{\citenamefont{Molnar and Voloshin}(2003)}]{Molnar:2003ff}
\bibinfo{author}{\bibfnamefont{D.}~\bibnamefont{Molnar}} \bibnamefont{and}
  \bibinfo{author}{\bibfnamefont{S.~A.} \bibnamefont{Voloshin}},
  \bibinfo{journal}{Phys. Rev. Lett.} \textbf{\bibinfo{volume}{91}},
  \bibinfo{pages}{092301} (\bibinfo{year}{2003}), \eprint{nucl-th/0302014}.

\bibitem[{\citenamefont{Ravagli and Rapp}(2007)}]{Ravagli:2007xx}
\bibinfo{author}{\bibfnamefont{L.}~\bibnamefont{Ravagli}} \bibnamefont{and}
  \bibinfo{author}{\bibfnamefont{R.}~\bibnamefont{Rapp}},
  \bibinfo{journal}{Phys. Lett. B} \textbf{\bibinfo{volume}{655}},
  \bibinfo{pages}{126} (\bibinfo{year}{2007}), \eprint{0705.0021}.

\bibitem[{\citenamefont{Ravagli et~al.}(2009)\citenamefont{Ravagli, van Hees,
  and Rapp}}]{Ravagli:2008rt}
\bibinfo{author}{\bibfnamefont{L.}~\bibnamefont{Ravagli}},
  \bibinfo{author}{\bibfnamefont{H.}~\bibnamefont{van Hees}}, \bibnamefont{and}
  \bibinfo{author}{\bibfnamefont{R.}~\bibnamefont{Rapp}},
  \bibinfo{journal}{Phys. Rev. C} \textbf{\bibinfo{volume}{79}},
  \bibinfo{pages}{064902} (\bibinfo{year}{2009}), \eprint{0806.2055}.

\bibitem[{\citenamefont{Cassing and Bratkovskaya}(2009)}]{Cassing:2009vt}
\bibinfo{author}{\bibfnamefont{W.}~\bibnamefont{Cassing}} \bibnamefont{and}
  \bibinfo{author}{\bibfnamefont{E.}~\bibnamefont{Bratkovskaya}},
  \bibinfo{journal}{Nucl.Phys.} \textbf{\bibinfo{volume}{A831}},
  \bibinfo{pages}{215} (\bibinfo{year}{2009}), \eprint{0907.5331}.

\bibitem[{\citenamefont{Greco et~al.}(2004)\citenamefont{Greco, Ko, and
  Rapp}}]{Greco:2003vf}
\bibinfo{author}{\bibfnamefont{V.}~\bibnamefont{Greco}},
  \bibinfo{author}{\bibfnamefont{C.~M.} \bibnamefont{Ko}}, \bibnamefont{and}
  \bibinfo{author}{\bibfnamefont{R.}~\bibnamefont{Rapp}},
  \bibinfo{journal}{Phys. Lett.} \textbf{\bibinfo{volume}{B595}},
  \bibinfo{pages}{202} (\bibinfo{year}{2004}), \eprint{nucl-th/0312100}.

\bibitem[{\citenamefont{Hwang}(2002)}]{Hwang:2001th}
\bibinfo{author}{\bibfnamefont{C.-W.} \bibnamefont{Hwang}},
  \bibinfo{journal}{Eur. Phys. J. C} \textbf{\bibinfo{volume}{23}},
  \bibinfo{pages}{585} (\bibinfo{year}{2002}).

\bibitem[{\citenamefont{Albertus et~al.}(2004)\citenamefont{Albertus, Amaro,
  Hernandez, and Nieves}}]{Albertus:2003sx}
\bibinfo{author}{\bibfnamefont{C.}~\bibnamefont{Albertus}},
  \bibinfo{author}{\bibfnamefont{J.~E.} \bibnamefont{Amaro}},
  \bibinfo{author}{\bibfnamefont{E.}~\bibnamefont{Hernandez}},
  \bibnamefont{and} \bibinfo{author}{\bibfnamefont{J.}~\bibnamefont{Nieves}},
  \bibinfo{journal}{Nucl. Phys. A} \textbf{\bibinfo{volume}{740}},
  \bibinfo{pages}{333} (\bibinfo{year}{2004}).

\bibitem[{\citenamefont{Peterson et~al.}(1983)\citenamefont{Peterson,
  Schlatter, Schmitt, and Zerwas}}]{Peterson:1982ak}
\bibinfo{author}{\bibfnamefont{C.}~\bibnamefont{Peterson}},
  \bibinfo{author}{\bibfnamefont{D.}~\bibnamefont{Schlatter}},
  \bibinfo{author}{\bibfnamefont{I.}~\bibnamefont{Schmitt}}, \bibnamefont{and}
  \bibinfo{author}{\bibfnamefont{P.~M.} \bibnamefont{Zerwas}},
  \bibinfo{journal}{Phys. Rev.} \textbf{\bibinfo{volume}{D27}},
  \bibinfo{pages}{105} (\bibinfo{year}{1983}).

\bibitem[{\citenamefont{Cacciari et~al.}(2005)\citenamefont{Cacciari, Nason,
  and Vogt}}]{Cacciari:2005rk}
\bibinfo{author}{\bibfnamefont{M.}~\bibnamefont{Cacciari}},
  \bibinfo{author}{\bibfnamefont{P.}~\bibnamefont{Nason}}, \bibnamefont{and}
  \bibinfo{author}{\bibfnamefont{R.}~\bibnamefont{Vogt}},
  \bibinfo{journal}{Phys. Rev. Lett.} \textbf{\bibinfo{volume}{95}},
  \bibinfo{pages}{122001} (\bibinfo{year}{2005}), \eprint{hep-ph/0502203}.

\bibitem[{\citenamefont{Cacciari et~al.}(2012)\citenamefont{Cacciari, Frixione,
  Houdeau, Mangano, Nason, and Ridolfi}}]{Cacciari:2012ny}
\bibinfo{author}{\bibfnamefont{M.}~\bibnamefont{Cacciari}},
  \bibinfo{author}{\bibfnamefont{S.}~\bibnamefont{Frixione}},
  \bibinfo{author}{\bibfnamefont{N.}~\bibnamefont{Houdeau}},
  \bibinfo{author}{\bibfnamefont{M.~L.} \bibnamefont{Mangano}},
  \bibinfo{author}{\bibfnamefont{P.}~\bibnamefont{Nason}}, \bibnamefont{and}
  \bibinfo{author}{\bibfnamefont{G.}~\bibnamefont{Ridolfi}},
  \bibinfo{journal}{JHEP} \textbf{\bibinfo{volume}{10}}, \bibinfo{pages}{137}
  (\bibinfo{year}{2012}), \eprint{1205.6344}.

\bibitem[{\citenamefont{Biro et~al.}(1995)\citenamefont{Biro, Levai, and
  Zimanyi}}]{Biro:1994mp}
\bibinfo{author}{\bibfnamefont{T.}~\bibnamefont{Biro}},
  \bibinfo{author}{\bibfnamefont{P.}~\bibnamefont{Levai}}, \bibnamefont{and}
  \bibinfo{author}{\bibfnamefont{J.}~\bibnamefont{Zimanyi}},
  \bibinfo{journal}{Phys. Lett. B} \textbf{\bibinfo{volume}{347}},
  \bibinfo{pages}{6} (\bibinfo{year}{1995}).

\bibitem[{\citenamefont{Scardina et~al.}(2010)\citenamefont{Scardina, Di~Toro,
  and Greco}}]{Scardina:2010zz}
\bibinfo{author}{\bibfnamefont{F.}~\bibnamefont{Scardina}},
  \bibinfo{author}{\bibfnamefont{M.}~\bibnamefont{Di~Toro}}, \bibnamefont{and}
  \bibinfo{author}{\bibfnamefont{V.}~\bibnamefont{Greco}},
  \bibinfo{journal}{Phys. Rev. C} \textbf{\bibinfo{volume}{82}},
  \bibinfo{pages}{054901} (\bibinfo{year}{2010}), \eprint{1009.1261}.

\bibitem[{\citenamefont{Liu et~al.}(2007)\citenamefont{Liu, Ko, and
  Zhang}}]{Liu:2006sf}
\bibinfo{author}{\bibfnamefont{W.}~\bibnamefont{Liu}},
  \bibinfo{author}{\bibfnamefont{C.}~\bibnamefont{Ko}}, \bibnamefont{and}
  \bibinfo{author}{\bibfnamefont{B.}~\bibnamefont{Zhang}},
  \bibinfo{journal}{Phys. Rev. C} \textbf{\bibinfo{volume}{75}},
  \bibinfo{pages}{051901} (\bibinfo{year}{2007}), \eprint{nucl-th/0607047}.

\bibitem[{\citenamefont{Olive et~al.}(2014)}]{Agashe:2014kda}
\bibinfo{author}{\bibfnamefont{K.}~\bibnamefont{Olive}} \bibnamefont{et~al.}
  (\bibinfo{collaboration}{Particle Data Group}), \bibinfo{journal}{Chin. Phys.
  C} \textbf{\bibinfo{volume}{38}}, \bibinfo{pages}{090001}
  (\bibinfo{year}{2014}).

\bibitem[{\citenamefont{Acharya et~al.}(2019{\natexlab{b}})}]{Acharya:2019mgn}
\bibinfo{author}{\bibfnamefont{S.}~\bibnamefont{Acharya}} \bibnamefont{et~al.}
  (\bibinfo{collaboration}{ALICE}), \bibinfo{journal}{Eur. Phys. J. C}
  \textbf{\bibinfo{volume}{79}}, \bibinfo{pages}{388}
  (\bibinfo{year}{2019}{\natexlab{b}}), \eprint{1901.07979}.



\end{thebibliography}
\end{document}